\documentclass[pra,showpacs,preprintnumbers,amsmath,amssymb]{revtex4}
\usepackage[cp1250]{inputenc}
\usepackage{amssymb}
\usepackage{amsmath}
\usepackage{graphicx}
\begin{document}

\title[Charge transfer with target excitation]
{Comment on "Four-body charge transfer processes in proton–helium
collisions" }

\author{S Houamer$^1$\footnote
             {s$\underline{~}$\ houamer@univ-setif.dz} and Yu V Popov$^2$\footnote
             {popov@srd.sinp.msu.ru}}

\address{$^1$Laboratoire de Physique Quantique et Syst$\grave{\rm e}$mes Dynamiques,
D$\acute{\rm e}$partement de Physique, Facult$\acute{\rm e}$ des
Sciences, Universit$\acute{\rm e}$ Ferhat Abbas, S$\acute{\rm
e}$tif, Algeria}
\address{$^2$ Skobeltsyn Institute of Nuclear Physics, Lomonosov Moscow State University, Moscow, Russia}
%\mailto{s$\underline{~}$\ houamer@univ-setif.dz},
%\mailto{popov@srd.sinp.msu.ru}

\begin{abstract}
We found, within the plane-wave first Born approximation (PWFBA),
that the proton-helium fully differential cross section (FDCS) for
transfer excitation agrees well with the experimental one at the
proton energy $E_p=300$ keV and small scattering angles both in
shape and in magnitude. This result is in a contradiction with
that obtained in~\cite{Madison12}.
\end{abstract}

 \vspace{2pc} \noindent{\it Keywords}: transfer
excitation, proton impact, helium atom

%\submitto{\JPB}

%Uncomment for PACS numbers title message
%\pacs{00.00, 20.00, 42.10}
% Keywords required only for MST, PB, PMB, PM, JOA, JOB?
%\vspace{2pc}
%\noindent{\it Keywords}: Article preparation, IOP journals
% Uncomment for Submitted to journal title message
%\submitto{\JPA}
% Comment out if separate title page not required
\maketitle

In Ref.~\cite{Madison12}, the authors consider the charge transfer
with target excitation reaction (TTE) p+He$\to$ H+He$^{+*}$ at
various proton energies and small scattering angles (few mrad).
They calculated FDCS in comparison with the absolute experimental
values~\cite{Frankfurt}. In turn, the experimentalists claimed
that they are able to resolve separately contributions from the
ground and excited He$^+$ states. The theorists used a specially
developed 9D code for calculation of principal matrix elements. In
such a way, they presented FBA (we call it PWFBA to emphasize that
both the proton and the hydrogen motion is described by plane
waves in the center-of-mass reference frame) calculations of FDCS
(see Fig. 4 in \cite{Madison12}). They found that a fitting factor
of about $v_p^4$ is necessary in order to compare theory and
experiment.

The 9D integral, which describes the FBA matrix element for TTE in
configurational space, can be reduced to a 3D one in momentum
space, if plane waves are used for the proton and hydrogen
center-of-mass motion and a trial helium wave function is of the
configuration-interaction (CI) type. Results of such calculations
are presented in Fig. 1 We take into account the following He$^+$
excited states: $2s,2p,3s,3p,3d$. Three CI trial wave functions
are used for the helium ground state: Roothaan-Hartree-Fock (RHF)
\cite{RHF} with a binding energy of -2.8617 a.u.,
Silverman-Platas-Matsen (SPM) \cite{SPM} with a binding energy of
-2.8952 a.u. and Mitroy \cite{Mitroy} with a binding energy of
-2.9031 a.u. The binding-energy value reflects a quality of the He
wave function. We find practically ideal agreement between
calculations using a well-correlated Mitroy wave function and
experiment at small angles ($\theta\leq 0.4$ mrad). For larger
angles the agreement is less impressive, but in this particular
case we expect a notable contribution from the second-Born term
(we saw an analogous behavior for the $1s-1s$ transition
\cite{Hong}). We also find a splitting of the FDCS curves, which
is characteristic of the TTE and TI reactions (see, for instance,
\cite{Salim10}).

In Fig. 2, we present, for the sake of comparison, our
calculations and those of Chowdhury and coworkers
\cite{Madison12}. We include also the FDCS calculations with
adding the leading $1s$ state to the sum. We remark that
contribution of excited states is about 5\% of the $1s-1s$
transition, which is quite expected for $E_p=300$ keV.

In conclusion, we think that the 9D code requires a careful
revision. We frequently encounter the effect that a choice of the
integration scheme and computation grid strongly influences the
results. Particularly it is relevant to calculations of tiny
values against the background of large values. We wish every luck
to the authors of the 9D code in overcoming the problems, as such
a code is highly desired for various DWBA calculations of the
capture and direct excitation/ionization processes induced by a
heavy-projectile impact. In this case, we can test different
distortion factors, as well as different target ground-state wave
functions.

\section*{Acknowledgements}
This work is partially supported by the Russian Foundation for
Basic Research (RFBR), Grant No.11-01-00523-a. We are very
thankful to D. Madison for stimulating discussions and U.
Chowdhury  for providing us with their results and experimental
points.

\newpage

\begin{figure}
\includegraphics[width=16cm]{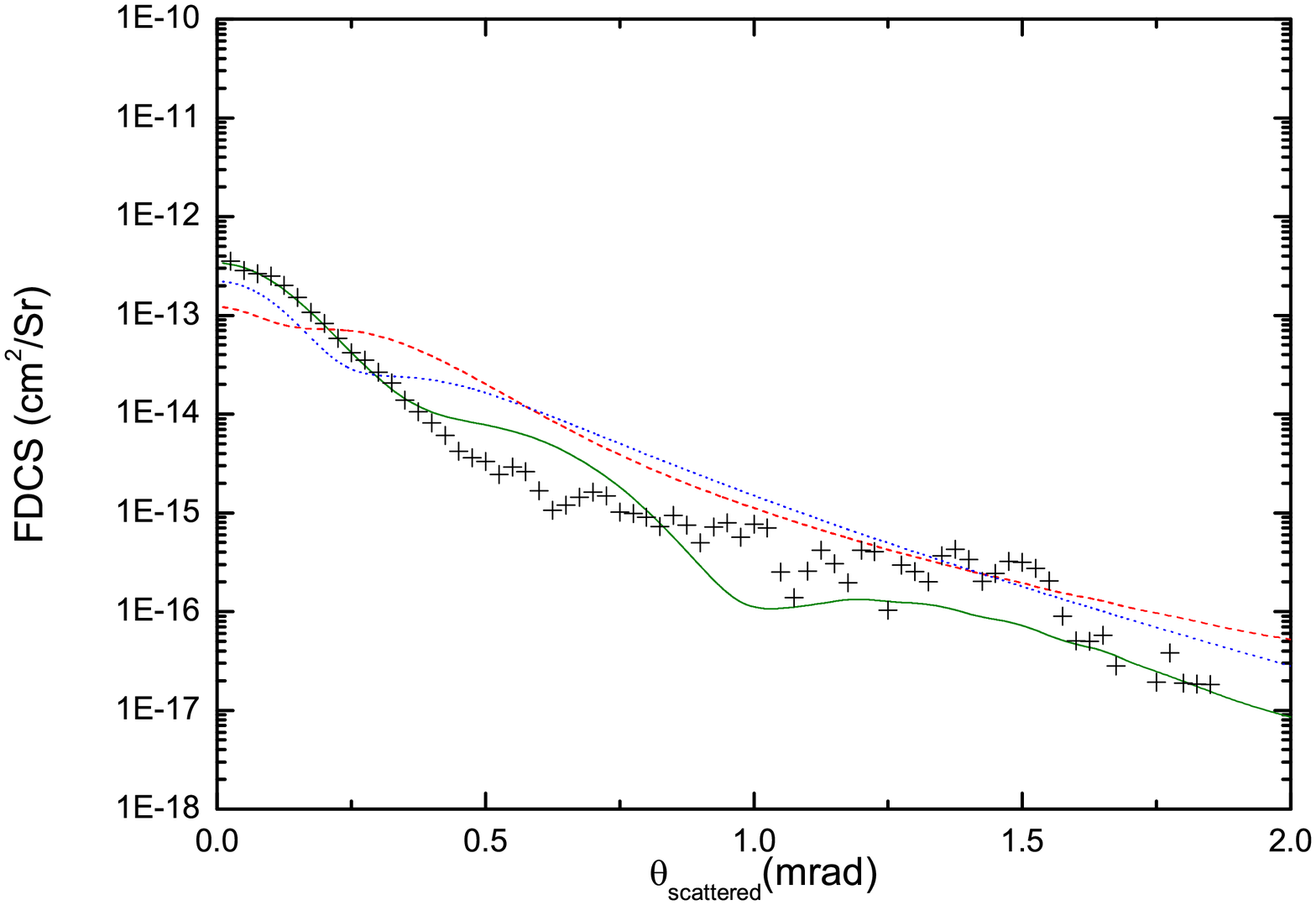}
\caption{FDCS as a function of the projectile scattering angle
$\theta$ in the lab system for three different target wave
functions. Solid line (green) corresponds to Mitroy, dashed (red)
to RHF, and dotted (blue) to SPM. Experimental values are due to
Sch\"offler \cite{Frankfurt}.}
\end{figure}
\begin{figure}
\includegraphics[width=16cm]{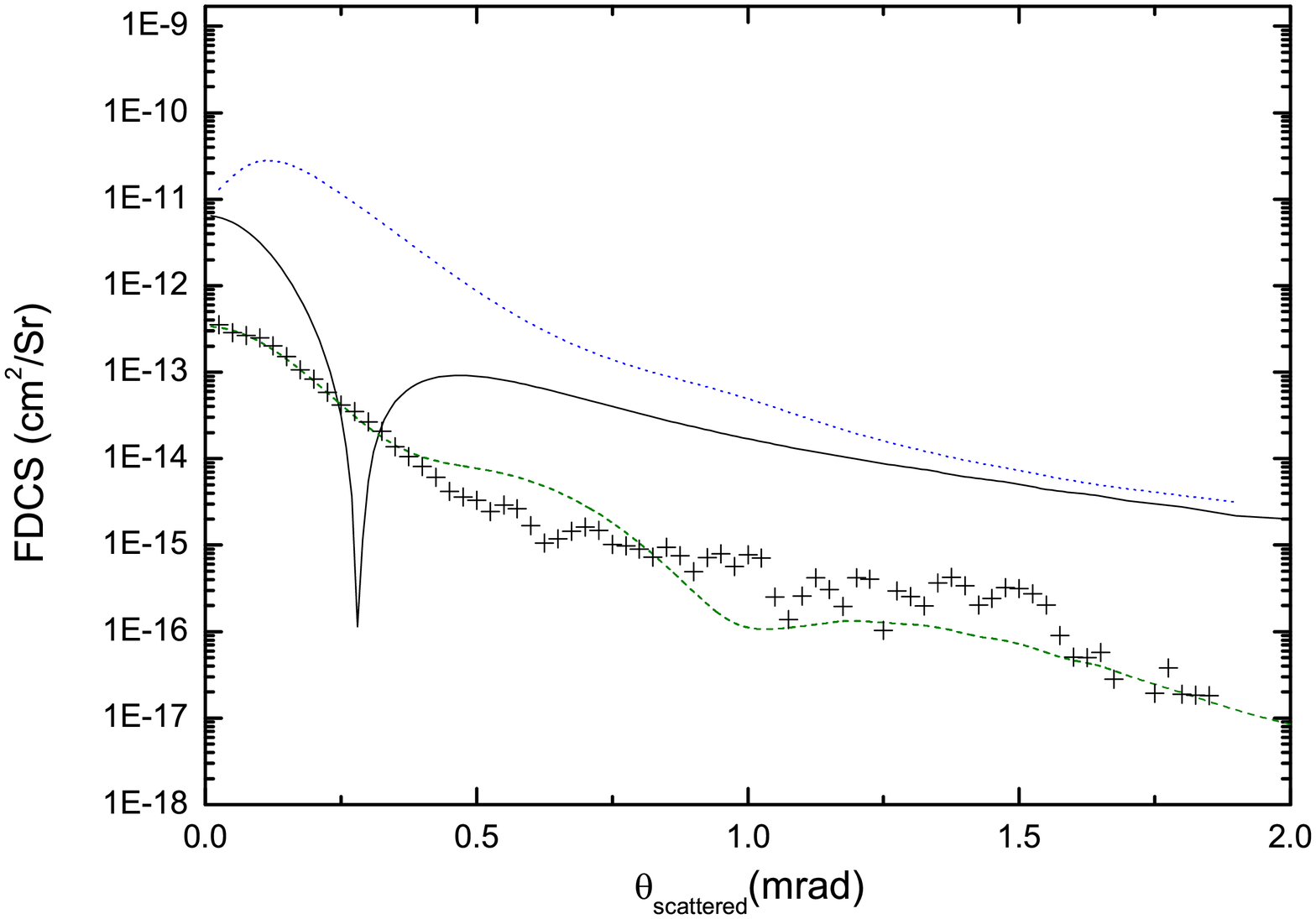}
\caption{FDCS as a function of the projectile scattering angle
$\theta$ in the lab system. Solid (black) line corresponds to
inclusion of the $1s$ state, dashed (green) to Mitroy for
He$^{+*}$ , and dotted (blue) to calculations from
\cite{Madison12}. The experiment is due to Sch\"offler
\cite{Frankfurt}.}

\end{figure}


\bigskip

\section*{References}
\begin{thebibliography}{99}

%\bibitem{SokTerSynRad68}
%    Sokolov A A and Ternov I M 1968 {\it Synchrotron radiation} (Oxford: Pergamon Press)

\bibitem{Madison12}
Chowdhury U, Harris A L, Peacher J L  and Madison D H, 2012 {\it
J. Phys B: At. Mol. Opt. Phys.} {\bf 45} 035203

\bibitem{Frankfurt} Sch\"offler M S, 2006 PhD Thesis University of Frankfurt am
Main

\bibitem{RHF} Enrico Clementi and Carla Roetti, 1974 {\it Atomic Data and
Nuclear Data Tables} \textbf{14} 177

\bibitem{SPM} Silverman J N, Platas O and Matsen F A, 1960 {\it J. Chem. Phys.} {\bf 32}
1402.

\bibitem{Mitroy} Mitroy J, McCarthy I E  and Weigold E, 1985 {\it J. Phys. B:
At. Mol. Phys.} {\bf 18} 4149

\bibitem{Hong} Hong-Keun Kim, Sch\"offler M S, Houamer S, Chuluunbaatar O,
Titze J N, Schmidt L Ph H, Jahnke T, Schmidt-B\"ocking H, Galstyan
A, Popov Yu V and D\"orner R 2012 {\it Phys. Rev. A.} \textbf{85}
022707.

\bibitem{Salim10} Houamer S, Popov Yu V and Dal Cappello C
2010 {\it Phys. Rev. A} \textbf{81} 032703.

\end{thebibliography}
\end{document}